\documentclass[12pt,letterpaper]{article}

\usepackage[intlimits]{amsmath}
\usepackage{amssymb}
\usepackage{bbm}
\usepackage{cite}
\usepackage{subfigure}
\usepackage[scale={.76,.76}]{geometry}
\usepackage{caption}
\newlength{\textlength}
\newlength{\overlinelength}
\newcommand{\ol}[2][.625]{%
   \settowidth{\textlength}{\ensuremath{#2}}%
   \setlength{\overlinelength}{3pt}%
   \addtolength{\overlinelength}{0.4\textlength}%
   \makebox[\textlength][s]{\ensuremath{#2}}%
   \hspace{-.5\textlength}\hspace{-\overlinelength}\hspace{#1\overlinelength}
   \overline{%
      \makebox[\overlinelength][c]{%
         \vphantom{\ensuremath{#2}}
      }
   }
   \hspace{-#1\overlinelength}\hspace{.5\textlength}
}
\newcommand{\loongrightarrow}{\ensuremath{\xrightarrow{\mspace{68mu}}}}

\numberwithin{equation}{section}
\numberwithin{table}{section}
\allowdisplaybreaks

\begin{document}
  \date{}

  \title{ 
    \vspace{1cm}
    {\bfseries The Potential Fate of Local Model Building\\[8mm]}
    \author{Christoph~L\"udeling, Hans~Peter~Nilles, Claudia~Christine~Stephan\\[2mm]
      {\ttfamily \normalsize luedeling, nilles, cstephan @th.physik.uni-bonn.de}\\[3mm]
      {\large\itshape Bethe Center for Theoretical Physics}\\{\normalsize\itshape and}\\
        {\large\itshape Physikalisches Institut der Universit\"at Bonn,}\\
      {\normalsize\itshape Nu\ss allee 12, 53115 Bonn, Germany}}}
  \maketitle
  
  \thispagestyle{empty}
  
  \vspace{0.7cm}
  \begin{abstract}
    \noindent
    We analyse local models at the point of $E_8$ in F-theory GUTs
    and identify exactly two models with potentially realistic
    properties concerning proton stability and a suitable pattern
    of quark and lepton masses. To this end we identify a matter parity at the local point. A globally consistent
    ultraviolet completion turns out to be problematic for both models. It is impossible to embed the models in a
    semilocal scheme via the  spectral cover approach. This seems to severely limit the predictive power of local model
    building and to indicate that the full string theory might give us valuable hints for particle physics model building.
  \end{abstract}
  
  \newpage
  \tableofcontents

  \section{Introduction\label{sec:intro}}
    Exceptional groups might play a crucial role in the incorporation 
    of grand  unified theories (GUTs) in the framework of string theory. 
    This is obvious in the $E_8 \times E_8$ heterotic theory \cite{Gross:1985fr}, and 
    more recently $E_8$ has been considered in F-theory \cite{Vafa:1996xn} as well. 
    Of course, $E_8$ is not an acceptable symmetry for a GUT in d=4 space 
    time dimensions. The breakdown of $E_8$ and further properties of the 
    d=4 theory depend crucially on the process of compactification. 
    Such a picture has been analysed in the framework of 
    the ``heterotic braneworld'' \cite{Forste:2004ie} where it led to the concept of 
    ``local grand unification''\footnote{Note that, in contrast to ``local models'', the concept of ``local grand
      unification'' refers to {\em global} models where the gauge symmetry exhibits a locally nontrivial profile in the
      compactified extra dimensions.}
    \cite{Kobayashi:2004ya,Buchmuller:2005jr,Lebedev:2006kn}. The  
    geography of localised  
    matter fields and the  nontrivial profile of gauge symmetries in 
    compactified space allowed a successful incorporation of 
    grand unification within string theory, thus providing 
    a consistent ultraviolet completion (for a review see \cite{Nilles:2008gq}). 
    F-theory leads to a qualitatively similar picture through 
    intersecting branes and localised fields in extra dimensions. 
    In contrast to the heterotic case, F-theory model building so far 
    has generally relied on a bottom-up approach (see, however, \cite{Blumenhagen:2009yv}) analysing mostly the
    vicinity of special (local) points in compactified space,
    while ignoring the constraints from global consistency 
    of the underlying theory (for a review see \cite{Weigand:2010wm}). 
    
    The most attractive and predictive set-up concerns 
    ``The point of $E_8$'' \cite{Heckman:2009mn}, where various branes cross at a point 
    with local symmetry enhancement to $E_8$. 
    This local picture has been analysed with respect to particle 
    physics model building
    \cite{Donagi:2008ca,Beasley:2008dc,Hayashi:2009ge,Marsano:2009gv,Dudas:2009hu,King:2010mq,Pawelczyk:2010xh,Choi:2010gx}
    with the goal to 
    obtain the minimal supersymmetric extension of the standard model (MSSM). 
    The present paper is an attempt to investigate 
    the predictive power of such a local model and see which 
    properties of particle physics could find a local explanation. 
    In a second step we shall then apply constraints from 
    global consistency and its implications for the local construction.
    
    From all the properties of GUTs or the MSSM it is the superpotential 
    that should find its explanation through the properties of the local model. The 
    superpotential is relevant for Yukawa couplings (thus quark and 
    lepton masses) as well as potential  dangerous terms that might 
    lead to fast proton decay. In the MSSM such terms are forbidden 
    by a symmetry like matter parity $P_M$ \cite{Dimopoulos:1981dw} (or generalisations 
    thereof \cite{Ibanez:1991pr,Dreiner:2005rd}). We would therefore like to address the following question: 
    \begin{itemize}     
    \item[*] Does the local model allow for proton stability with correct 
      quark and lepton masses?
    \end{itemize}

    Correct Yukawa couplings should provide the top quark mass at 
    the trilinear level and might explain other quark and lepton masses 
    (if zero at the trilinear level) at higher order via a variant of 
    the Froggatt--Nielsen mechanism \cite{Froggatt:1978nt}.
    
    Our analysis is performed in the framework of an $SU(5)$ GUT as 
    defined in Ref.~\cite{Dudas:2009hu}.  Higher unified groups like $SO(10)$ or $E_6$ 
    will lead to more restrictive model building and will not be 
    considered here. In Section~\ref{sec:pointofe8} we shall present the setup of 
    local models in detail. We identify the various curves 
    that could support matter multiplets in $\boldsymbol{10}$ and $\boldsymbol{\ol[.5]{5}}$
    representations  of $SU(5)$, as well as $\boldsymbol{5}$ and $\boldsymbol{\ol[.5]{5}}$ 
    representations for Higgs fields and singlets. We then display the
    general form of the superpotential that might be generated
    in such a scheme.
    
    Section~\ref{sec:unique} analyses the prospects for realistic model 
    building. We assume a trilinear coupling for the top quark 
    and identify candidates for matter parity $P_M$. The scheme 
    turns out to be quite restrictive: Only two candidates for 
    $P_M$ are allowed. We then identify the curves that carry the 
    various matter and Higgs fields and check whether they
    allow for all quark and lepton masses and are consistent 
    with $P_M$ and proton stability. A scan of all possibilities 
    then leads to two surviving models, exactly one for 
    each of the two choices of $P_M$. This is a remarkable result: 
    The local $E_8$ point is rather predictive.  
    
    It remains to be seen, however, whether the two local 
    models can be incorporated within a globally consistent scheme. 
    To construct the local models we had to  make some assumptions about fluxes that 
    determine the chirality of matter fields and split the Higgs multiplets. 
    Such assumptions are restricted from global considerations. 
    In a first step towards a global realisation we consider a 
    semilocal framework where information 
    about the  8-dimensional (but not yet the full 12-dimensional)
    GUT surface is included. To perform the analysis we use the 
    so-called spectral cover approach \cite{Friedman:1997yq,Donagi:2009ra} as the (so far) only  
    available tool for this discussion (see, however, \cite{Cecotti:2010bp}). 
    A summary of spectral cover results is given in Section~\ref{sec:semilocal} 
    followed by a scan of the local models, with particular 
    emphasis on the two successful models of Section~\ref{sec:unique}. 
    The result is quite striking: Both models are ruled out 
    since there is no consistent extension to a global model. 
    They fail already at the level of the semilocal completion.
    This suggests that realistic models of particle physics 
    need to have some ``nonlocal ingredients'' that are not
    captured by the local point. 
    Thus the predictions of local models might not be 
    trustworthy in absence of a consistent global completion. 
    A thorough discussion of F-theory ``predictions'' as 
    well as conclusion and outlook will be given in Section~\ref{sec:conclusion}.


  \section{Local Models, Operators and Matter Parity\label{sec:pointofe8}}
    F-theory GUT models\cite{Beasley:2008dc,Donagi:2008ca} are a generalisation of type~IIB 
    intersecting branes which allow for exceptional local symmetry groups, and thus in particular for a top quark
    Yukawa coupling, which requires a local $E_6$ enhancement. Similar to intersecting branes, one usually considers local
    models, in which one concentrates on the branes carrying the GUT symmetry, or curves and points within it. This
    ``bottom-up'' approach has the usual advantage that there is much more freedom in model building and
    one can ignore many of the problems of a global construction. The obvious disadvantage is that one does not know whether
    there is a global completion. Furthermore, some issues such as moduli stabilisation can only be tackled in a global 
    model. 

    In   Section~\ref{subsec:localF} we will introduce the framework of local F-theory models. Since this is
    by now well-known, we will be rather brief, for a detailed review see e.g.~\cite{Weigand:2010wm}.
    After that,  we introduce the relevant operators -- Yukawa couplings and proton decay -- and the
    matter parity we want to impose.

    \subsection{Local Models\label{subsec:localF}}
      Global F-theory models describe general vacua of type~IIB string theory in terms of an elliptically
      fibred Calabi--Yau fourfold $X$, where seven-branes are indicated by the degeneration locus of the
      elliptic fibration.  The general idea of local F-theory GUT models is to decouple the bulk of $X$ and
      focus instead on a seven-brane on a submanifold $S$ where the GUT symmetry, which will be
      $G_\text{GUT}=SU(5)$ in our case, is localised. The intersections with other branes are visible as enhancements
      of the local symmetry group to $G_\Sigma\supset G_\text{GUT}$ on curves $\Sigma$ of complex codimension
      one. On these matter curves, there are localised hypermultiplets, the representation of which can
      be inferred from the decomposition of the adjoint of $G_\Sigma$. For $G_\text{GUT}=SU(5)$, we get
      matter in the $\boldsymbol{5}$ and $\boldsymbol{10}$ representations from enhancements to $SU(6)$
      and $SO(10)$, respectively.

      Furthermore, the matter curves can intersect in points, and on these intersections the gauge group
      enhances even further to $G_P$. There will be localised Yukawa couplings from the cubic
      interaction of the adjoint of $G_P$. For an $SU(5)$ GUT, the up-type and down-type Yukawa couplings
      require pointwise enhancements to $E_6$ and $SO(12)$, respectively. 

      Since the surface $S$ locally carries gauge groups larger than $G_\text{GUT}$, there is a
      different perspective: One can (at least locally) think of the worldvolume theory on $S$ as a gauge
      theory with larger gauge group which is broken by a position-dependent vacuum expectation value
      (VEV) for an adjoint Higgs 
      field. Since we want at least an $E_6$ enhancement, the largest possible gauge group is $E_8$,
      which contains $G_\text{GUT}=SU(5)$ and its commutant, $E_8\supset SU(5)\times SU(5)_\perp$. This
      is then broken to $SU(5)\times U(1)^4$ by the Higgs field. The extra $U(1)$'s correspond to the
      transverse branes and are generically broken in F-theory\footnote{This is a global issue that cannot be analysed in a local
        model\cite{Grimm:2010ez}.}, but remain as global selection rules 
      for the Lagrangean.  

      The matter curves are now the
      loci where certain components of the Higgs field vanish, such that the unbroken group is enhanced.
      To see what representations are localised where, we note the decomposition of the
      $\boldsymbol{248}$ of $E_8$:
      \begin{align}\label{eq:248decomposition}
        \begin{split}
          E_8&\loongrightarrow  SU(5)\times SU(5)_\perp\\
          \boldsymbol{248}&\loongrightarrow  \left(\boldsymbol{24},\boldsymbol{1}\right) \oplus
          \left(\boldsymbol{1},\boldsymbol{24}\right) \oplus
          \left(\boldsymbol{10},\boldsymbol{5}\right)\oplus\left(\ol[.5]{\boldsymbol{5}},\boldsymbol{10}\right)
          \oplus \left(\ol[.5]{\boldsymbol{10}},\ol[.5]{\boldsymbol{5}}\right) \oplus
          \left(\boldsymbol{5},\ol[.5]{\boldsymbol{10}}\right)
        \end{split}
      \end{align}
      We consider a diagonalisable Higgs field $\Phi$ and introduce a basis $e_i$, $i=1,\dotsc,5$, for the
      $\boldsymbol{5}$ of $SU(5)_\perp$, such that $\Phi e_i=t_i e_i$. The Higgs eigenvalues $t_i$ have
      to satisfy the tracelessness condition  
      \begin{align}\label{eq:t-t-traceless}
        \sum_{i=1}^5 t_i=0\,.
      \end{align}
      The $\boldsymbol{10}$ and the non-Cartan elements of the adjoint are spanned by the
      antisymmetric products $e_i\wedge e_j$ and by $e_i\wedge e_j^*$, respectively, where $i\neq j$. These
      elements are again eigenvectors of $\Phi$ with eigenvalues $t_i+t_j$ and $t_i-t_j$. The $t_i$ vary
      over the GUT surface, and the matter is localised on their vanishing locus. From the
      decomposition~(\ref{eq:248decomposition}) we can see how the representations of $SU(5)$ and
      $SU(5)_\perp$ are paired up, and focusing now on the
      representations of the unbroken $SU(5)$, we find the following equations for the matter curves:
      \begin{align}\label{eq:t-reps}
        \boldsymbol{10}\!: & \; t_i=0\,,& \boldsymbol{5}\!: & -\left(t_i+t_j\right)=0\,, & \boldsymbol{1}\!: &
        \pm\left(t_i-t_j\right)=0\,. 
      \end{align}
      For the $\ol[.5]{\boldsymbol{10}}$ and $\ol[.5]{\boldsymbol{5}}$, there is an overall minus sign in the
      equation, because they correspond to conjugate representations of $SU(5)_\perp$. 

      The $t_i$ determine the geometry of a deformed $E_8$ singularity in terms of the Tate model
      \begin{align}\label{eq:tate}
        y^2&=x^3 + b_5 xy +b_4 x^2 w + b_3yw^2 +b_2 xw^3+ b_0 w^5\,.
      \end{align}
      The GUT surface $S$ is located at $w=0$, and the coefficients $b_k$ are the elementary symmetric
      polynomials in the $t_i$ of degree $k$.  We can rephrase the conditions~(\ref{eq:t-reps}) for the
      localisation of matter representations in terms of the $b_k$, which yields
      \begin{align}\label{eq:b-reps}
        \boldsymbol{10}\!: & \; b_5=0\,,& \boldsymbol{5}\!:\, & b_3^2 b_4-b_2b_3b_5+b_0 b_5^2=0\,.
      \end{align}
      Note that the $b_k$ do not fully specify the $t_i$, since the relation is
      nonlinear. In particular, this means there can be monodromies interchanging some of the $t_i$
      (since the $b_i$ are symmetric polynomials), which amounts to identifying matter curves, and
      effectively reducing the commutant $SU(5)_\perp$ and the number of $U(1)$'s that remain.

      If all matter curves meet at one single point, then all the $t_i=0$, and thus all $b_k=0$ except
      $b_0$. At this point the singularity is enhanced to $E_8$, as can be seen from Eq.~(\ref{eq:tate}), which becomes
      $y^2=x^3+w^5$. In the following we will assume
      that this is the case, and that all interactions come from this point of maximal enhancement. This
      has the advantage that the allowed interactions are determined purely by the quantum numbers, and
      there are no geometric suppression effects due to separations of Yukawa points.  It has also been
      argued in \cite{Heckman:2009mn} that this structure is favourable for incorporating proper masses and
      mixings in both the quark and lepton sector, including neutrino masses which we will ignore in this work.

      If we had a global model at hand, we could determine the $b_k$ and the matter curves, hence also
      the $t_i$. However, in the local approach we focus on the surface $S$ and the point of $E_8$
      enhancement. We will consider two degrees of locality: In the strictly {\em local} point of view, adopted in
      Section~\ref{sec:unique}, we
      consider only the single point of $E_8$, such that we are dealing with a purely four-dimensional theory. This in
      particular implies that we can freely choose the monodromy group and the
      chirality of the zero modes on the matter curves (recall that matter comes in hypermultiplets, so
      that the chiral spectrum will be determined by some index theorem involving fluxes on the matter
      curves), and we will further assume that we can use a globally trivial, but locally nontrivial hypercharge
      flux to achieve doublet-triplet splitting for the Higgs~$\boldsymbol{5}$'s without introducing
      exotics on other curves. In Section~\ref{sec:semilocal} we will then consider a {\em semilocal} approach, in
      which we use spectral cover techniques to see whether this can actually be realised. At this
      level, we  still assume a decoupled bulk, but take a more global look at the GUT surface $S$, i.e.\ we consider an
      eight-dimensional model. In particular, we realise the monodromy by a suitable choice of spectral cover, and find
      correlations between the homology classes of various matter curves. This implies that the restrictions of the
      hypercharge flux to the curves are correlated as well. Hence, it will turn out that the 
      assumption of easy doublet-triplet splitting is not satisfied, which will lead to problems in the
      spectrum\cite{Marsano:2009wr,Dudas:2009hu}.

    \subsection{The Good, the Bad, the Parity\label{subsec:operators}}
      Our aim in the following sections will be to find a $SU(5)$ GUT model, i.e.\ an assignment of
      fields to matter curves, which satisfies some basic phenomenological requirements. In particular,
      we demand masses for up- and down-type quarks and 
      proton stability. To this end we will look for a matter parity $P_M\subset SU(5)_\perp$ which
      forbids many operators leading to proton decay. We assume that all operators which are allowed by
      gauge symmetry and matter parity are generated with order one coefficients.

      For the mass terms, some (good) operators of the form
      \begin{align}
        \boldsymbol{1}_a \dotsm\boldsymbol{1}_b  \boldsymbol{5}_{H_u} \boldsymbol{10}_M
        \boldsymbol{10}_M& \,,& \boldsymbol{1}_a \dotsm\boldsymbol{1}_b \ol{\boldsymbol{5}}_{H_d}
        \ol{\boldsymbol{5}}_M \boldsymbol{10}_M& 
      \end{align}
      should be allowed, so that VEVs for the singlets $\boldsymbol{1}_a$ lead to mass terms for all
      quarks and leptons. However, we require the
      $\boldsymbol{10}\,\boldsymbol{10}\,\boldsymbol{5}$ Yukawa coupling for the third generation at
      tree level to ensure a heavy top quark.

      \medskip

      The relevant baryon and lepton number violating (bad) operators are \cite{Dudas:2009hu}
      \begin{align}
        \begin{split}\label{eq:Wblv}
          W_{\not B,\not L}&=\beta_i \ol[.5]{\boldsymbol{5}}_M^i \boldsymbol{5}_{H_u} + \lambda_{ijk}
          \ol[.5]{\boldsymbol{5}}_M^i \ol[.5]{\boldsymbol{5}}_M^j \boldsymbol{10}_M^i
          +W^1_{ijkl}\boldsymbol{10}_M^i \boldsymbol{10}_M^j \boldsymbol{10}_M^k
          \ol[.5]{\boldsymbol{5}}_M^l\\  
          & \quad +  W^2_{ijk}\boldsymbol{10}_M^i \boldsymbol{10}_M^j \boldsymbol{10}_M^k
          \ol[.5]{\boldsymbol{5}}_{H_d} + W^3_{ij}\ol{\boldsymbol{5}}_M^i \ol{\boldsymbol{5}}_M^j
          \boldsymbol{5}_{H_u}\boldsymbol{5}_{H_u} +  W^4_{i}\ol[.5]{\boldsymbol{5}}_M^i
          \ol[.5]{\boldsymbol{5}}_{H_d}  \boldsymbol{5}_{H_u}\boldsymbol{5}_{H_u}
        \end{split}
      \end{align}
      from the superpotential and 
      \begin{align}
        \begin{split}\label{eq:Kblv}
          K_{\not B,\not L}&= K^1_{ijk}\boldsymbol{10}_M^i \boldsymbol{10}_M^j \boldsymbol{5}_M^k +
          K^2_i \ol[.5]{\boldsymbol{5}}_{H_u} \ol[.5]{\boldsymbol{5}}_{H_d}\boldsymbol{10}_M^i
        \end{split}
      \end{align}
      from the K\"ahler potential. Again the coefficients may contain singlet VEVs. 

      Matter
      parity\cite{Dimopoulos:1981dw,Ibanez:1991pr} is a $\mathbbm{Z}_2$ symmetry 
      under which the matter fields are odd and the Higgs fields are even:
      \begin{align}
        \begin{array}{c|c|c}
          & \ol[.5]{\boldsymbol{5}}_M^i, \boldsymbol{10}_M^i & \boldsymbol{5}_{H_u},
          \ol[.5]{\boldsymbol{5}}_{H_d} \\\hline
          P_M & -1 & +1
        \end{array}
      \end{align}
      $P_M$ forbids all operators in Eqns.~(\ref{eq:Wblv}) and~(\ref{eq:Kblv}) except $W^1$ and $W^3$. The
      $W^3$ operator leads to neutrino masses via the Weinberg operator $LLHH$. This might still be
      allowed if it is generated at higher order in the singlets. $W^1$, on the other hand, is very
      tightly constrained and must be very strongly suppressed.

      We will later see that when splitting the Higgs curves, some of the $\mathbf{10}$ multiplets will be split as
      well. So one could wonder whether it is possible that the $W^1$ operator is present at the $SU(5)$ 
      level, but the split will be such that for the SM multiplets, there are no dangerous terms.
      To see that this is not the case, we write out $W^1$ in terms of SM representations,
      \begin{align}
        \begin{split}
          W^1_{ijkl}\mathbf{10}_M^i\mathbf{10}_M^j\mathbf{10}_M^k\ol[.5]{\mathbf{5}}_M^l&= W^1_{ijkl}\ol
          e^i\ol u^jQ^kL^l+ W^1_{ijkl}\ol e^i\ol u^j\ol u^k\ol d^l \\
          &\quad\mspace{80mu}+ W^1_{ijkl}\ol u^iQ^jQ^k\ol d^l+
          W^1_{ijkl}Q^iQ^jQ^kL^l \,.
        \end{split}
      \end{align} 
      The first term, for example, is only absent for all permutations of $i$, $j$ and $k$ when the
      $\ol[.5]{\mathbf{5}}_M^l$ is split such that the chirality of $L^l$ is zero. For the second term,
      however, the same argument applies for $\ol d^l$. Since this reasoning works for all $l$ and
      because we must require that at least the sum of the chiralities over all $l$ is minus three for
      both $L$ and $\ol d$, we conclude that $W^1$ must be absent at the $SU(5)$ level.


  \section{Local Models with Matter Parity\label{sec:unique}}
    This section outlines the steps that lead to the two local models presented in this work. To specify a
    concrete model one has to choose the monodromy, define the matter parity $P_M$, assign matter and
    Higgs fields to the curves and select a set of singlets that get a VEV. This list could give the 
    impression that there is a lot of freedom in the attempt of building a local or semilocal model
    which incorporates matter parity. However, we can drastically reduce the number of options by imposing a few
    reasonable phenomenological requirements.

    Our requirements are:
    \begin{enumerate}\addtocounter{enumi}{-1}
      \item A matter parity $P_M\subset SU(5)_\perp$,
      \item a heavy top quark, i.e.\ a tree-level rank-one up-type Yukawa coupling involving the third generation
        $\boldsymbol{10}$ curve, 
      \item absence of dimension-five proton decay via the $W^1$ operator,
      \item masses for all quarks and leptons after switching on singlet VEVs.
    \end{enumerate}

    We will start with general arguments that must be valid in the local
    as well as in the semilocal framework by showing in Section~\ref{subsec:parity} that there are only
    two possible definitions of matter parity and essentially one choice of monodromy group. In the subsequent
    sections we will first consider case~I, demonstrating in Section~\ref{subsec:fields} that the way to
    assign matter fields to curves is very restricted when 
    requiring the absence of proton decay. In Section~\ref{subsec:higgs} we discuss the possibilities to
    choose the down-type Higgs curve and we will see that this leads in fact to a unique local model for case~I
    whose phenomenology will be examined in Section~\ref{subsec:flavor}. Afterwards, in
    Section~\ref{sec:caseII} a similar analysis will be performed for case~II, where there is the possibility of
    enlarging the monodromy group, but without introducing qualitatively new features.

    \subsection{Matter Parity and Monodromy\label{subsec:parity}}
      We would first like to motivate the choice of the monodromy group and the definition of matter
      parity. We require that there is a tree-level coupling $\mathbf{10}_M
      {\mathbf{10}}_M\mathbf{5}_{H_u}$ that leads to a heavy top quark. Since the top quark and the
      anti-top quark both are in the same $\mathbf{10}$ representation, this can only be achieved provided
      the $\mathbf{10}_M$'s participating in the mass term are the same. Keeping in mind that the up-type
      Higgs has a charge of the form $-t_i-t_j$, the $\mathbf{10}_M$'s must have charges $t_i$ and $t_j$
      for the mass term to be gauge invariant. Since $t_i\neq t_j$, see
      Eq.~(\ref{eq:t-reps}), this would imply that the $\mathbf{10}$'s are different and thus to allow the
      top-anti-top quark coupling at least a $\mathbbm{Z}_2$ monodromy is required. We
      choose its action to be $t_1\leftrightarrow t_2$, so that the top quark generation is assigned to the
      corresponding curve $\mathbf{10}_1$ given in terms of the weight representation by $\{t_1,t_2\}$.
      Then there exists a tree-level up-type Yukawa coupling that leads to a heavy top quark provided we
      also fix the up-type Higgs curve $\mathbf{5}_{H_u}$ to be the curve with the charge $-t_1-t_2$. 
      
      Our matter parity emerges from the $SU(5)_\perp$ and therefore must be defined in terms of the
      $t_i$. Each $t_i$ can either contribute a factor of $+1$ or $-1$ and thus a formula for $P_M$ can
      be written in full generality as
      $P_M=(-1)^{\alpha_1t_1+\alpha_2t_2+\alpha_3t_3+\alpha_4t_4+\alpha_5t_5}$, where $\alpha_i$ takes the
      values 1 or 2. Other values for the $\alpha$'s would not give anything new because it is a
      $\mathbbm{Z}_2$ symmetry. Note that the up-type Yukawa coupling will always be allowed by matter
      parity because the requirement of gauge invariance alone leads to $P_M(\text{up-type Yukawa
        coupling})=(-1)^0$ since the $t$'s cancel and this conclusion persists no matter how many singlets
      are inserted.  
      
      The requirement that the down-type Yukawa couplings $\mathbf{10}_M\ol[.5] {\mathbf{5}}_M\ol[.5]
      {\mathbf{5}}_{H_d}$ should be allowed does give a constraint on the matter parity definition though:
      Given that the $\mathbf{10}_M$ contributes a factor of $t_i$ and the $\ol[.5]{\mathbf{5}}$'s contribute
      $t_j+t_k$ and $t_l+t_m$, all of which have a positive sign, this operator can, in contrast to the
      up-type Yukawa coupling, only be gauge invariant if all $t$'s are different, so that we get
      $t_1+t_2+t_3+t_4+t_5$, which is zero according to Eq.~(\ref{eq:t-t-traceless}). At the same time the desired
      down-type Yukawa coupling must have matter parity $+1$ to be permitted, which can only be achieved
      provided the number of $t$'s with a prefactor of 2 in the matter parity definition is odd. Note
      that this fact remains true with any number of singlet insertions because the singlets all have
      charge assignments of the form $t_i-t_j$ and thus do not change the number of $t$'s in the operator.
      When setting all five $\alpha$'s to 2 there will not be a single field left that we could identify
      with matter since matter must have $P_M=-1$. 
      One option is to set only a single $\alpha$ to 2, which we choose to be $\alpha_5$:
      \begin{align}
        \text{ Case I}:\quad P_M=(-1)^{t_1+t_2+t_3+t_4+2t_5}\,. 
      \end{align}
      This model will be analysed in Section~\ref{sec:caseI}.
      Finally, having three prefactors of 2 in the matter parity
      definition forces us to build a model where three generations come from a single $\mathbf{10}_M$ curve, namely
      the top curve. We will examine the model corresponding to the matter parity 
      \begin{align}
        \text{Case II}:\quad P_M=(-1)^{t_1+t_2+2t_3+2t_4+2t_5}
      \end{align}
      in Section~\ref{sec:caseII}. 

      Now let us come back to the choice of the monodromy group. In case I, for $P_M$ to be well-defined, $t_5$ must
      not be related to any other $t$. So there are only two options left which maintain the chance of
      building a model where the three generations of the standard model emerge from at least two curves.
      The first one is to enlarge the monodromy group such that another $t$ lies in the same orbit as
      $t_1$ and $t_2$ and the second one is to additionally relate $t_3$ and $t_4$ by a $\mathbbm Z_2$
      monodromy. Both ideas are to be discarded because they are accompanied by the occurrence of the
      operator $W^1_{ijkl}\mathbf{10}_M^i\mathbf{10}_M^j\mathbf{10}_M^k\ol[.5]{\mathbf{5}}_M^l$, which leads
      to proton decay, see Section~\ref{subsec:operators}. The origin of this issue is that a gauge invariant $W^1$
      operator also needs a sum over all five distinct $t$'s, as it is the case for the down-type Yukawa
      couplings. More precisely, since the three $\mathbf{10}$'s in $W^1$ each add a $t$ with a prefactor
      of one, for an allowed $W^1$ the $\ol[.5]{\mathbf{5}}$ must provide the remaining two $t$'s, in
      particular a factor of $t_5$ that cannot get in via the $\mathbf{10}_M$'s. A $\ol[.5]{\mathbf{5}}$
      always has a charge assignment $t_i+t_j$. So for it to have $P_M=-1$ and from our definition of
      matter parity it is evident that one of these $t$'s must in fact be the $t_5$. Now the only chance
      to avoid $W^1$ is to not identify one of the odd matter parity $\mathbf{10}$'s with SM matter. Since
      already with the $\mathbbm{Z}_2$ monodromy there are only 3 odd parity curves left, it is evident
      that the monodromy group must not be enlarged any further.

      In case II, $t_3$, $t_4$ and $t_5$ appear symmetrically in the definition of $P_M$. Here it is
      possible to mutually relate them by an arbitrary monodromy, but we will see later that this does not
      affect the phenomenology of the resulting model and therefore we will leave it at the monodromy
      relating $t_1$ and $t_2$.

    \subsection{Matter Parity Case I\label{sec:caseI}}
      \subsubsection{Matter Curves and Singlet VEVs\label{subsec:fields}}
        Having fixed the monodromy group and a formula for matter parity,
        \begin{align*}
          P_M=(-1)^{t_1+t_2+t_3+t_4+2t_5}\,,
        \end{align*}
        we proceed with the
        discussion which fields to assign to the different curves. The aim of this selection is to
        prevent the appearance of baryon and lepton number violating 
        operators. We have collected the curves, their charges and matter parities in Table~\ref{tabfields}. As already
        mentioned in the previous section, all dimension three, four and five 
        baryon and lepton number violating operators apart from
        $W^3_{ij}\ol[.5]{\mathbf{5}}_M^{i}\ol[.5]{\mathbf{5}}_M^{j}\mathbf{5}_{H_u}\mathbf{5}_{H_u}$ and
        $W^1_{ijkl}\mathbf{10}_M^i\mathbf{10}_M^j\mathbf{10}_M^k\ol[.5]{\mathbf{5}}_M^l$ are forbidden by
        matter parity. Since the two up-type Higgs curves in $W^3$ contribute a factor $-2t_1-2t_2$ to the
        operator, $W^3$ is absent at tree level because these charges cannot be canceled by adding the
        $t$'s for the other two curves. On the other hand, $W^1$ will appear at tree level, as we have
        argued above, unless we do not assign SM matter to some of the odd matter parity curves.
  
        From Eq.~(\ref{equW1}), which lists all gauge invariant combinations involving the
        $\mathbf{10}$'s and $\mathbf{5}$'s with $P_M=-1$, one can see that it is possible to evade $W^1$
        when not assigning SM fields to the curves $\mathbf{10}_2$ and $\mathbf{5}_5$ because if these two
        curves do not carry SM fields, there is no operator left that contains SM fields only\footnote{One
          could also pick $\mathbf{10}_3$ and $\mathbf{5}_6$ instead but this choice just amounts to a
          relabeling.}.  
        \begin{align}
          \begin{split}\label{equW1}
            \mathbf{10}_1 \mathbf{10}_1 \mathbf{10}_2 \ol[.5]{\mathbf{5}}_6 \\
            \mathbf{10}_1 \mathbf{10}_1 \mathbf{10}_3 \ol[.5]{\mathbf{5}}_5 \\
            \mathbf{10}_1 \mathbf{10}_2 \mathbf{10}_3 \ol[.5]{\mathbf{5}}_3
          \end{split}
        \end{align}
        
        Hence, the $\mathbf{10}$ and $\mathbf{5}$ curves carrying SM matter are fixed to be $\mathbf{10}_1$
        and $\mathbf{10}_3$ as well as $\mathbf{5}_3$ and $\mathbf{5}_6$, respectively, because one can see
        from Table~\ref{tabfields} that these are the only remaining fields with odd matter parity. Because
        of this we are forced to build a model where three fields emerge from two curves.
        
        Addressing the even matter parity $\mathbf{5}$ curves, there are four possibilities to assign the
        down-type Higgs to one of the Higgs-like curves. The choice must be made such that after turning on
        VEVs for selected singlets the down quarks become massive without reintroducing operators that lead
        to proton decay. We do not give VEVs to odd matter parity singlets because this would break matter
        parity and reintroduce the successfully eliminated baryon and lepton number violating operators.
        $W^1$ will be generated immediately when VEVs are given to the singlets $\mathbf{1}_{1}$ or
        $\mathbf{1}_{4}$ because the matter $\mathbf{10}$'s have the charges $t_1,t_2$ and $t_4$ and the
        matter $\mathbf{5}$'s have the charges $\{t_1+t_5,t_2+t_5\}$ and $t_4+t_5$. $\mathbf{1}_{1}$ and
        $\mathbf{1}_{4}$ have the charges $\pm \{t_1-t_3,t_2-t_3\}$ and $\pm(t_3-t_4)$ and thus both contain
        a $t_3$ which is needed for $W^1$ to be gauge invariant. Therefore they must not get a VEV. Summing
        up, the aim is to select a down-type Higgs curve that gives masses to the down-type quarks using
        only VEVs for the singlets $\mathbf{1}_{2}$ and $\mathbf{1}_{7}$. The assignment of fields to the matter curves is
        summarised in the last column of Table~\ref{tabfields}.
        \begin{table}[t]
          \centering
          $\begin{array}{c|ccc}
            & \text{\textbf{Charge}}&\text{\textbf{Matter Parity}} &\text{\textbf{Assigned Fields}} \\[1mm]
            \hline
            &&&\\
            \text{\textbf{10 Curves}}&&&\\
            \mathbf{10}_1   & \{t_1,t_2\}&-& \text{$\boldsymbol{10}_\text{top}$, possibly more} \\
            \mathbf{10}_2   & t_3 &-& \text{no SM matter} \\
            \mathbf{10}_3 & t_4 &-& \text{possible SM matter} \\
            \mathbf{10}_4   & t_5 &+& \text{no SM matter}  \\
            &&& \\
            \text{\textbf{5 Curves}} & & &\\
            \mathbf{5}_{H_u}   & -t_1-t_2 &+& \text{up-type Higgs}\\
            \mathbf{5}_{1}   & \{-t_1-t_3,-t_2-t_3\}&+& \text{Higgs-like}\\
            \mathbf{5}_{2}  & \{-t_1-t_4,-t_2-t_4\}&+&\text{Higgs-like}\\
            \mathbf{5}_{3}   & \{-t_1-t_5,-t_2-t_5\}&-& \text{possible SM matter}\\
            \mathbf{5}_{4}   & -t_3-t_4 &+& \text{Higgs-like}\\
            \mathbf{5}_{5}   & -t_3-t_5&-& \text{no SM matter}\\
            \mathbf{5}_{6}   & -t_4-t_5&-& \text{possible SM matter}\\
            &&&  \\
            \text{\textbf{Singlets}} & & &\\
            \mathbf{1}_{1}  &\pm\{t_1-t_3,t_2-t_3\}&+&  \text{no VEV}\\
            \mathbf{1}_{2}   &\pm\{t_1-t_4,t_2-t_4\}&+& \text{VEV possible}\\
            \mathbf{1}_{3}   &\pm\{t_1-t_5,t_2-t_5\}&-& \text{no VEV}\\
            \mathbf{1}_{4}   &\pm(t_3-t_4)&+& \text{no VEV}\\
            \mathbf{1}_{5}   &\pm(t_3-t_5)&-& \text{no VEV}\\
            \mathbf{1}_{6}   &\pm(t_4-t_5)&-& \text{no VEV}\\
            \mathbf{1}_{7}   &\{t_1-t_2,t_2-t_1\}&+& \text{VEV possible}\\
          \end{array}$
          \caption{List of curves, their charges and their matter parity values with the field
            assignment for case I.\label{tabfields}} 
        \end{table}

      \subsubsection{Higgs Curves\label{subsec:higgs}}
        In this section we will see that, when working in the purely local framework, requiring no proton
        decay and down-type masses at the same time singles out a unique model for case~I. The main assumption is that
        the chiralities of the curves can be chosen at will while simultaneously the Higgs curves can
        be split correctly, that is, only the Higgs doublets remains light. In particular, we assume that all
        $\mathbf{10}$ and $\mathbf{5}$ curves apart from $\mathbf{10}_1$, $\mathbf{10}_3$,
        $\mathbf{5}_{H_u}$, $\ol[.5]{\mathbf{5}}_{H_d}$, $\ol[.5]{\mathbf{5}}_{3}$ and $\ol[.5]{\mathbf{5}}_{6}$
        appear as vector-like pairs and can be given a high-scale mass. In Section~\ref{sec:semilocal} this
        point will be analysed in more detail. 
        
        The previous section tells us that, since the SM matter curves are fixed, the next important question
        is to which of the four possible even matter parity $\mathbf{5}$ curves $\mathbf{5}_{H_u}$,
        $\mathbf{5}_1$, $\mathbf{5}_2$ and $\mathbf{5}_4$ the down-type Higgs field is assigned.
        Table~\ref{tabdownhiggses} lists all gauge invariant tree-level down-type Yukawa couplings for the
        different choices of the down-type Higgs curve.  
        
        \begin{table}[ht]
          \centering
          $\begin{array}{c|c}
            \text{\textbf{Down-type Higgs curve}} &\text{\textbf{Gauge invariant couplings}}\\[1mm]
            \hline
            &\\
            \ol[.5]{\mathbf{5}}_{H_u}& \text{---}\\[3mm]
            \hline
            & \\
            \ol[.5]{\mathbf{5}}_{1} &\ol[.5]{\mathbf{5}}_{H_d}\mathbf{10}_1\ol[.5]{\mathbf{5}}_6 \\
            & \ol[.5]{\mathbf{5}}_{H_d}\mathbf{10}_3\ol[.5]{\mathbf{5}}_3\\[3mm]
            \hline
            &  \\
            \ol[.5]{\mathbf{5}}_{2} & \text{---}\\[3mm]
            \hline
            &  \\
            \ol[.5]{\mathbf{5}}_{4}&\ol[.5]{\mathbf{5}}_{H_d}\mathbf{10}_1\ol[.5]{\mathbf{5}}_3\\
          \end{array}$
          \caption{Gauge invariant down-type Yukawa terms for all possible choices of down-type Higgs
            curves for case~I.\label{tabdownhiggses}} 
        \end{table}
        Taking the down-type Higgs curve to be $\ol[.5]{\mathbf{5}}_{1}$ leads to a rank-two Yukawa matrix at
        tree level resulting in two heavy and one light generations, which is phenomenologically problematic.
        One can check that a particular split of the curves reduces the rank of the matrix to one or zero,
        but since with the spectral cover formalism it is not possible to realise the $\ol[.5]{\mathbf{5}}_{1}$
        as the down-type Higgs curve anyway, as we will see later, we dismiss this option here and relegate 
        more details to Appendix~\ref{app:higgs} . 

        Consider next the choice of $\ol[.5]{\mathbf{5}}_{H_u}$ or $\ol[.5]{\mathbf{5}}_2$ with charges 
        $\{t_1+t_2\}$ and $\{t_1+t_4,t_2+t_4\}$, respectively. It is easy to see that in both cases there
        cannot be any down-type masses because the matter $\mathbf{10}$'s, $\mathbf{10}_1$ and
        $\mathbf{10}_3$, have charges $\{t_1,t_2\}$ and $t_4$, while the matter $\ol[.5]{\mathbf{5}}$'s,
        $\ol[.5]{\mathbf{5}}_3$ and $\ol[.5]{\mathbf{5}}_6$, have charges $\{t_1+t_5,t_2+t_5\}$ and $t_4+t_5$, and
        the only two singlets that we are allowed to give a VEV to, as discussed in the previous section,
        are $\mathbf{1}_2$ and $\mathbf{1}_7$ with charges $\pm\{t_1-t_4,t_2-t_4\}$ and
        $\{t_1-t_2,t_2-t_1\}$. For the down-type mass term to be gauge invariant, a sum over all five
        distinct $t$'s is needed and with this choice it is obvious that the sum will always lack a $t_3$
        factor. Thus, the possibilities to select $\ol[.5]{\mathbf{5}}_{H_u}$ or $\ol[.5]{\mathbf{5}}_2$ as the
        down-type Higgs curve are excluded. Note that this conclusion holds even if the curves are
        eventually split.

        This ultimately fixes the down-type Higgs curve to be $\ol[.5]{\mathbf{5}}_4$
        \begin{align}
          \ol[.5]{\mathbf{5}}_{H_d}= \ol[.5]{\mathbf{5}}_4\,,
        \end{align} 
        which is favoured anyway because it leads to a rank-one down-type Yukawa matrix at tree level.
        Demanding that the quark which gets the large mass is the bottom quark then amounts to assigning
        the bottom quark generation to the curve $\ol[.5]{\mathbf{5}}_3$ and the light generations to the curve
        $\ol[.5]{\mathbf{5}}_6$,
        \begin{align}
          \ol[.5]{\mathbf{5}}_\text{bottom}=\ol[.5]{\mathbf{5}}_3\,, \quad
          \ol[.5]{\mathbf{5}}_\text{down/strange}=\ol[.5]{\mathbf{5}}_6 \,.
        \end{align}

        Recapitulating, the requirements of no proton decay, a heavy top quark and a down-type Yukawa matrix
        which has rank zero or one have guided us to a unique model. The next step is to explore its
        phenomenology. 

      \subsubsection{Masses and Mixings\label{subsec:flavor}}
        Let us now examine the Yukawa textures and Cabibbo--Kobayashi--Maskawa (CKM) matrix in the
        model that was just specified to see whether reasonable mass hierarchies and mixings in the quark
        sector can be achieved.

        Choosing $\mathbf{10}_1$ to carry only the top generation and $\mathbf{10}_3$ the light generations,
        calculating the up-type Yukawa matrix including insertions of the singlets $\mathbf{1}_2$ and
        $\mathbf{1}_7$ leads, at leading order, to the following result.  
        \begin{align}
          Y_{ij}^u&\sim \begin{pmatrix} \epsilon^2 & \epsilon^2 & \epsilon \\ \epsilon^2 & \epsilon^2 & \epsilon
            \\ \epsilon&\epsilon&1 \end{pmatrix}\,, 
        \end{align}
        where
        \begin{align}
          \epsilon=\frac{\langle X_2\rangle}{M_*}\,.
        \end{align}
        Here $\langle X_2\rangle$ is the VEV for the field assigned to the curve $\mathbf{1}_2$, which is
        suppressed by the winding scale $M_*$ which is also the GUT scale for local
        models\cite{Conlon:2009qa}. 
        The VEV for $\mathbf{1}_7$ will only appear at higher order and
        so can be ignored at this level.

        It is important to note that there are order-one prefactors in front of each entry, which depend on
        the geometry of the different curves and also come from integrating out heavy states in the case of
        elements with singlet insertions. Within this framework these prefactors cannot be determined,
        but at this point we are not interested in the details of the matrices.  Instead, we would like to
        see if they show acceptable patterns.  

        For the down-type Yukawa matrix we get a similar result:
        \begin{align}
          Y_{ij}^d&\sim\begin{pmatrix} \epsilon^2 & \epsilon^2 & \epsilon \\ \epsilon^2 & \epsilon^2 & \epsilon
            \\ \epsilon&\epsilon&1 \end{pmatrix} \,.
        \end{align}
        Since these matrices are both approximately diagonal, we can use the simplified formulae for the
        mixing angles 
        in the CKM matrix \cite{Hall:1993ni}
        \begin{align}
          s_{ij}^\text{CKM}\simeq s_{ij}^d-s_{ij}^u\,, 
        \end{align}
        where 
        \begin{align}
          s_{12}^{u/d}=\frac{Y_{12}^{u/d}}{Y_{22}^{u/d}}\,,\quad
          s_{13}^{u/d}=\frac{Y_{13}^{u/d}}{Y_{33}^{u/d}}\,,\quad s_{23}^{u/d}=\frac{Y_{23}^{u/d}}{Y_{33}^{u/d}}\,,
        \end{align}
        and we arrive at
        \begin{align}
          V_\text{CKM}&\sim \begin{pmatrix} 1 & 1 & \epsilon \\ 1 & 1 & \epsilon \\ \epsilon&\epsilon&1 \end{pmatrix}\,.
        \end{align}
        These results show that in our model where the Yukawa matrices and the CKM matrix are described by a
        single parameter $\langle X_2\rangle$, a mass can be given to all generations and in addition there
        is also some mixing. Neither the Yukawa matrices nor the CKM matrix match the SM data very
        well, but this was not expected because in our setup three generations come
        from only two curves and thus there will always be a certain degeneracy in the entries of the Yukawa
        matrices and thus also in the CKM matrix.

    \subsection{Matter Parity Case II\label{sec:caseII}}
      In this section we analyse the matter parity case II along the lines of the previous
      section: We will first clarify the possible field assignment and then discuss in Section~\ref{sec:shlonce2} the
      down-type Higgs sector and the resulting Yukawa textures. 
      
      The matter parity in case~II is defined as
      \begin{align*}
        P_M=(-1)^{t_1+t_2+2t_3+2t_4+2t_5}\,.
      \end{align*} 
      It will turn out that one has in general more freedom in
      this model because the $SU(5)_\perp$ charges split into two more or less decoupled groups,
      $t_\text{even}=\{t_3,t_4,t_5\}$ and $t_\text{odd}=\{t_1,t_2\}$, such that the possible Higgs $\boldsymbol{5}$ curves
      involve two $t_\text{even}$'s and the matter  $\boldsymbol{5}$ curves involve one $t_\text{even}$ and one
      $t_\text{odd}$. Furthermore, the positive matter parity singlets do not mix $t_\text{even}$ and $t_\text{odd}$.
      
      As in case~I, we find a basically unique model. There is only one matter $\boldsymbol{10}$ curve, and both Higgses
      are unique (up to a relabeling of the $t_\text{even}$). The only freedom is in the choice of matter
      $\boldsymbol{\ol[.5]{5}}$ curves: One can choose one, two or three curves for the three generations, or
      alternatively identify some of these by an extended monodromy. However, this will not give qualitatively new features.

      \subsubsection{Matter Assignment\label{sec:IImatter}}  
        In the case at hand the different curves have even and odd matter parity as displayed in Table~\ref{tab:caseIIpm}.
        \begin{table}[t]
          \centering
          $\begin{array}{c|ccc}
            & \text{\textbf{Charge}}&\text{\textbf{Matter Parity}} & \text{\textbf{Assigned Fields}}\\[1mm]
            \hline
            &&&\\
            \text{\textbf{10 Curves}}&&&\\
            \mathbf{10}_1  & \{t_1,t_2\}& -& \text{all families}\\
            \mathbf{10}_2  & t_3  & +& \text{no SM matter}\\
            \mathbf{10}_3  & t_4 & +&\text{no SM matter} \\
            \mathbf{10}_4  & t_5  & +& \text{no SM matter} \\
            &&& \\
            \text{\textbf{5 Curves}} & & &\\
            \mathbf{5}_{H_u}  & -t_1-t_2 & +& \text{up-type Higgs}\\
            \mathbf{5}_{1} & \{-t_1-t_3,-t_2-t_3\}& -& \text{possible SM matter}\\
            \mathbf{5}_{2} & \{-t_1-t_4,-t_2-t_4\}& -&  \text{possible SM matter}\\
            \mathbf{5}_{3} & \{-t_1-t_5,-t_2-t_5\}& -&  \text{possible SM matter}\\
            \mathbf{5}_{4} & -t_3-t_4 & +&  \text{Higgs-like}\\
            \mathbf{5}_{5} & -t_3-t_5& +&  \text{Higgs-like}\\
            \mathbf{5}_{6} & -t_4-t_5& +&  \text{Higgs-like}\\
            &&&  \\
            \text{\textbf{Singlets}} & & &\\
            \mathbf{1}_{1} &\pm\{t_1-t_3,t_2-t_3\}& -& \text{no VEV}\\
            \mathbf{1}_{2} &\pm\{t_1-t_4,t_2-t_4\}& -& \text{no VEV}\\
            \mathbf{1}_{3} &\pm\{t_1-t_5,t_2-t_5\}& -& \text{no VEV}\\
            \mathbf{1}_{4} &\pm(t_3-t_4)& +&  \text{VEV possible}\\
            \mathbf{1}_{5} &\pm(t_3-t_5)& +&  \text{VEV possible}\\
            \mathbf{1}_{6} &\pm(t_4-t_5)& +& \text{VEV possible}\\
            \mathbf{1}_{7} &\{t_1-t_2,t_2-t_1\}& +& \text{VEV possible}\\
          \end{array}$
          \caption{List of curves, their charges and their matter parity values with the field
            assignment for case II.\label{tab:caseIIpm}} 
        \end{table}
        Since there is only one odd matter parity $\mathbf{10}$ curve in case II, all three generations of
        SM fields that belong to the $\mathbf{10}$ representation have to be assigned to $\mathbf{10}_1$.  
        Note, however, that one can choose the matter in the $\ol[.5]{\mathbf 5}$ representation to come from one, two or three
        $\ol[.5]{\mathbf 5}$ curves.

        The situation with the $W^1$ operator has improved compared to case I: 
        Using $t_\text{even}$ and $t_\text{odd}$ defined above, the operator
        $\mathbf{10}_M\mathbf{10}_M\mathbf{10}_M\ol[.5]{\mathbf{5}}_M$ then has the charge 
        $4t_\text{odd}+t_\text{even}$. 
        Therefore, with this choice of matter parity and
        under the assumption that VEVs are given only to matter parity even singlets, $W^1$ cannot be
        generated no matter which and how many singlets are inserted. This statement is also completely
        independent of the assignment of fields to the curves, which is only constrained by the matter
        parity and shown in Table~\ref{tab:caseIIpm}.

      \subsubsection{Higgs Assignment and Flavour\label{sec:shlonce2}}
        Since there is only one $\boldsymbol{10}$ matter curve, there is only one up-type Yukawa coupling,
        $\boldsymbol{5}_{H_u}\boldsymbol{10}_1 \boldsymbol{10}_1$, which at tree level leads to a Yukawa matrix 
        \begin{align}
          Y_{ij}^u&=
          \begin{pmatrix}
            0&0&0\\0&0&0\\0&0&1
          \end{pmatrix}\,.
        \end{align}
        We can get singlet contributions from e.g.\ the VEV of $\boldsymbol{1}_7$, or various higher powers of other singlet
        VEVs. This gives a generic form of the Yukawa matrix as 
        \begin{align}
          Y_{ij}^u&\sim
          \begin{pmatrix}
            \epsilon &\epsilon &\epsilon \\
            \epsilon &\epsilon &\epsilon \\
            \epsilon &\epsilon &1
          \end{pmatrix}\,.
        \end{align}
        Here the $\epsilon$'s again involve order-one coefficients.

        Turning to the down-type Yukawa couplings, there are four Higgs-like $\ol[.5]{\mathbf 5}$ curves. Having the
        down-type Higgs on the same curve as 
        the up-type Higgs, there will be no generation of down-type masses at any level as can be seen from
        the charges: The reason for the absence is the same as for the $W^1$ operator. We get for the
        coupling $\mathbf{10}_M\ol[.5]{\mathbf{5}}_M\ol[.5]{\mathbf{5}}_{H_u}$ again $4t_\text{odd}+t_\text{even}$ which, as
        already stated, can never be made gauge invariant by singlet insertions. All other choices for the
        down-type Higgs curve are equivalent, given that this model is invariant under permutations of $t_3$, $t_4$ and
        $t_5$ and 
        the three remaining Higgs-like curves and possible $\ol[.5]{\mathbf 5}$ matter curves only involve these
        charges. So let 
        us choose $\ol[.5]{\mathbf 5}_{H_d}=\ol[.5]{\mathbf 5}_{4}$. In order to have a tree-level coupling of the
        $\mathbf{10}_1$ curve to any down-type quark, we have to assign matter to the curve
        $\mathbf{5}_{3}$. It is also possible to start with a down-type Yukawa matrix of rank zero at tree
        level and generate all masses through singlet insertions. The allowed couplings to lowest order in
        the singlets between the down-type Higgs curve and the candidate SM matter $\ol[.5]{\mathbf 5}$ curves
        are: 
        \begin{align}
          \mathbf{10}_1\ol[.5]{\mathbf{5}}_3\ol[.5]{\mathbf{5}}_{H_d},\quad
          \mathbf{10}_1\ol[.5]{\mathbf{5}}_1\ol[.5]{\mathbf{5}}_{H_d}\mathbf{1}_5, \quad
          \mathbf{10}_1\ol[.5]{\mathbf{5}}_1\ol[.5]{\mathbf{5}}_{H_d}\mathbf{1}_4\mathbf{1}_6, \quad
          \mathbf{10}_1\ol[.5]{\mathbf{5}}_2\ol[.5]{\mathbf{5}}_{H_d}\mathbf{1}_6, \quad
          \mathbf{10}_1\ol[.5]{\mathbf{5}}_2\ol[.5]{\mathbf{5}}_{H_d}\mathbf{1}_4\mathbf{1}_5 
        \end{align}
        Depending on which singlets get a VEV and how the SM generations are assigned to the three
        $\ol[.5]{\mathbf 5}$ matter curves, one can arrive at different down-type Yukawa matrices. Starting
        with a rank-one matrix at tree level, one can assign the bottom quark generation to
        $\mathbf{5}_{3}$, and the first and second generation to $\mathbf{5}_{1}$ and $\mathbf{5}_{2}$,
        respectively. Switching on VEVs for $\mathbf{1}_5$ and $\mathbf{1}_4$, one arrives at a down-type
        Yukawa matrix, where some entries are generated with only one singlet insertion and others with two:        
        \begin{align}
          Y_{ij}^d&\sim\begin{pmatrix} \epsilon_5 & \epsilon_5 & \epsilon_5 \\ \epsilon_5\epsilon_4 &
            \epsilon_5\epsilon_4 & \epsilon_5\epsilon_4 
            \\ 0 &0 &1 \end{pmatrix} \,.
        \end{align}
        Switching on VEVs for $\mathbf{1}_5$ and $\mathbf{1}_6$, all entries involving $\ol d$ and $\ol b$
        are generated with only one singlet insertion:  
        \begin{align}
          Y_{ij}^d&\sim \begin{pmatrix} \epsilon_5 & \epsilon_5 & \epsilon_5 \\ \epsilon_6 & \epsilon_6 & \epsilon_6
            \\ 0&0&1 \end{pmatrix} \,.
        \end{align}
        Of course, there exists also the option to assign both light generations to one curve. Then, one
        singlet insertion is sufficient to get all couplings. 
        
        If one starts with a rank-zero matrix at tree level, all matter must be assigned to $\mathbf{5}_{1}$
        and $\mathbf{5}_{2}$.  Choosing $\mathbf{5}_{1}$ to carry the bottom generation and $\mathbf{5}_{2}$
        the other generations, one can for example realise the matrix 
        \begin{align}
          Y_{ij}^d&\sim \begin{pmatrix} \epsilon_5\epsilon_4 & \epsilon_5\epsilon_4 & \epsilon_5\epsilon_4 \\
            \epsilon_5\epsilon_4 & \epsilon_5\epsilon_4 & \epsilon_5\epsilon_4 
            \\ \epsilon_5&\epsilon_5&\epsilon_5 \end{pmatrix} 
        \end{align}
        or the same matrix with $\epsilon_5\epsilon_4$ replaced by $\epsilon_6$.

        Since they cannot be determined within this framework and we are not aiming at presenting a
        detailed discussion of flavour, we will now move on to the question whether the models presented in
        this section can be realised in a more global framework. 


  \section{Semilocal Realisation\label{sec:semilocal}}
    In this section we attempt to extend our local models to semilocal models, which would be the first
    step in searching for a global realisation. The important improvement of the semilocal framework,
    where the whole GUT surface $S_{GUT}$ is considered, is that the chiral spectrum of a model can be
    calculated explicitly. The chiralities of the curves are determined by the restrictions of two kinds
    of fluxes. One of them is turned on along the $U(1)$'s that remain after imposing the action of the
    monodromy group (i.e.\ along the transverse branes) and can only influence the chiralities of full GUT multiplets.
    The other flux is the hypercharge flux, which can split the multiplets. The latter, as opposed to the $U(1)$ flux,
    is confined to $S_{GUT}$ and thus, as we will see, leads to strong constraints on the spectrum of
    semilocal models. The aim of Section~\ref{subsec:coverresults} is, essentially following \cite{Dudas:2009hu}, to find a
    formula for the chiralities of the various fields in terms of the restrictions of the above fluxes. Therefore
    one needs to know the homology classes of the curves, which in turn can be calculated using the
    spectral cover approach. In Section~\ref{subsec:covercalculation} we will perform
    explicit 
    calculations using the spectral cover results and show that our local models, presented in
    Section~\ref{sec:unique}, unfortunately have no semilocal realisation. One should, however, keep
    in mind that, as was noted in~\cite{Cecotti:2010bp}, the spectral cover approach used here is not
    necessarily the most general framework. Thus, to exclude the local models once and for all, further
    studies seem to be necessary. 
    
    \subsection{Summary of Spectral Cover Results\label{subsec:coverresults}}
      In Section~\ref{sec:pointofe8} we discussed how the information about the monodromy group is
      contained in the deformation parameters $b_k$ appearing in Eq.~(\ref{eq:tate}). Another elegant
      way to handle the monodromy data is to work with the spectral cover. It is defined to be the surface
      given by the constraint 
      \begin{align}
        \label{equ:c10}
        C_{10}=b_0U^5+b_2V^2U^3+b_3V^3U^2+b_4V^4U+b_5V^5=0\,,
      \end{align}
      where $U$ and $V$ are homogeneous coordinates of the projective threefold
      \begin{align}
        X=\mathbb P(\mathcal O_{S_{GUT}}\oplus K_{S_{GUT}})\,.
      \end{align}
      $\mathcal O_{S_{GUT}}$ and $K_{S_{GUT}}$ denote the trivial and the canonical bundle on
      $S_{GUT}$, respectively. The monodromy is encoded in the factorisation of $C_{10}$ because the
      number of $U(1)$'s 
      that remain independent is in general one less than the number of factors of $C_{10}$. To visualise
      how this comes about one can picture all $\mathbf{10}$ curves to be one single $\mathbf{10}$ curve on
      that 
      five-sheeted spectral cover. The different sheets of that cover get connected by branch cuts so that
      it breaks into slices, each of which is associated to a factor of $C_{10}$ and
      corresponds to one $\mathbf{10}$ curve. Concretely, one can locally define a parameter $s=U/V$
      and the five roots of Eq.~(\ref{equ:c10}), written as a polynomial in $s$, will then correspond to the five
      $t_i$. It is now clear that in order to realise the $\mathbb Z_2$ monodromy,
      we need the following factorisation into four parts, where the curves $\mathbf{10}_1$ and
      $\mathbf{10}_2$ lift to a single curve on the spectral cover:
      \begin{align}
        \label{equ:split}
        C_{10} = \left(a_1 V^2 + a_2 V U + a_3 U^2 \right) \left(a_4 V + a_7 U\right) \left(a_5 V + a_8
          U\right) \left(a_6 V + a_9 U\right) = 0 
      \end{align}
      It is possible to solve for the $a_i$ in terms of the $b_i$ and to calculate their homology classes.
      Note that Eq.~(\ref{eq:tate}) implies that the
      $b_k$ are sections of line bundles with first Chern class $\eta-k c_1$, where 
      \begin{align}
        \eta=6c_1-t\,,\quad 
        t=-c_1(N)\,,\quad
        c_1=c_1(S_{GUT})\,, 
      \end{align}
      with $N$ the normal bundle of $S_{GUT}$. Since there are six $b$'s, but nine $a$'s, three
      bundles remain unspecified and are denoted $\chi_7$, $\chi_8$ and $\chi_9$ corresponding to
      $a_7$, $a_8$ and $a_9$. 
      
      The constraint one gets from 
      \begin{align}
        b_1 = t_1 + t_2 + t_3 + t_4 + t_5 = 0
      \end{align}
      implies
      \begin{align}
        a_2a_7a_8a_9 + a_3a_6a_7a_8 + a_3a_5a_7a_9 + a_3a_4a_8a_9 = 0
      \end{align}
      and is nontrivial. It can be solved by the ansatz
      \begin{align}
        \label{equ:solution}
        \begin{split}
          a_2 &= - c \left( a_6 a_7 a_8 + a_5 a_7 a_9 + a_4 a_8 a_9\right)\, \\
          a_3 &= c\; a_7 a_8 a_9\,,
        \end{split}
      \end{align}
      without inducing non-Kodaira singularities, as was shown in \cite{Dudas:2010zb}. This, however, does
      not need to be the only solution and might thus not constitute the most general one. Here, the
      homology class $[c]$ is introduced which is given by   
      \begin{align}
        [c] = \eta - 2 \tilde{\chi}\,,
      \end{align}
      where $\tilde{\chi}=\chi_7+\chi_8+\chi_9$. Table~\ref{tab:homclasses} summarises the Chern classes
      of the various bundles for all $a_i$. 
      \begin{table}
        \centering
        $\begin{array}{c|c}
          \hline
          \text{Section} & c_1(\text{Bundle})\\
          \hline
          a_1 & \eta - 2c_1 - \tilde{\chi} \\
          \hline
          a_2 & \eta - c_1 - \tilde{\chi} \\
          \hline
          a_3 & \eta - \tilde{\chi}\\
          \hline
          a_4 & -c_1 + \chi_7 \\
          \hline
          a_5 & -c_1 + \chi_8\\
          \hline
          a_6 & -c_1 + \chi_9\\
          \hline
          a_7 & \chi_7 \\
          \hline
          a_8 & \chi_8\\
          \hline
          a_9 & \chi_9\\
          \hline
        \end{array}$
        \caption{The first Chern classes of the line bundles corresponding to the $a_i$.\label{tab:homclasses}}
      \end{table}
      
      Next, we need to determine the matter curves in terms of the $a_i$. Eq.~(\ref{eq:t-reps})
      tells us that the $10$ curves are given by $t_i=0$ and this implies $b_5=t_1t_2t_3t_4t_5=0$, which
      is in turn the coefficient of $V^5$ and must also be equal to $a_1a_4a_5a_6$, as one can see from
      Eq.~(\ref{equ:split}). Therefore, one concludes that the $\mathbf{10}$ curves are given by
      $a_k=0$, where $k=1,4,5,6$.  
      
      In order to determine the equations for the $\mathbf 5$ curves, we need to plug~(\ref{equ:solution}) into the
      defining polynomial for the $\mathbf 5$ curves~(\ref{eq:b-reps}). This gives  
      \begin{align} 
        \begin{split}
          P_5 &= \left(a_5 a_7 + a_4 a_8 \right) \left(a_6 a_7 + a_4 a_9 \right) \left(a_6 a_8 + a_5 a_9 \right) \\
          & \quad \times \left( a_6 a_7 a_8 + a_5 a_7 a_9 + a_4 a_8 a_9\right) \\
          & \quad  \times \left( a_1 - c a_5 a_6 a_7 - c a_4 a_6 a_8 \right) \\
          & \quad \times  \left( a_1 - c a_5 a_6 a_7 - c a_4 a_5 a_9 \right)  \\
          & \quad \times  \left( a_1 - c a_4 a_6 a_8 - c a_4 a_5 a_9 \right)\,,
        \end{split}
      \end{align}
      and we arrive at Table~\ref{tab:a}, which displays the curves, their $SU(5)_\perp$ charges and their
      defining equations in terms of the $a_i$ as well as the resulting homology classes. 
      
      Now we can finally specify the spectrum in terms of the restrictions of the $U(1)$ fluxes and the
      hypercharge flux to the curves. If one denotes by the integers $M$ and $N_Y$ the restriction of the
      $U(1)$ fluxes and the hypercharge flux to a curve, then the $\mathbf 5$ curves get split in the
      following way: 
      \begin{align}
        \begin{split}
          n_{(3,1)_{-1/3}}-n_{(\bar 3,1)_{+1/3}}&=M_{\boldsymbol{5}}\,,\\
          n_{(1,2)_{+1/2}}-n_{(1,2)_{-1/2}}&=M_{\boldsymbol{5}}+N_Y\,,
          \label{equsplit5}\end{split}
      \end{align}
      and for the $\mathbf{10}$ curves we have
      \begin{align}
        \begin{split} 
          n_{(3,2)_{+1/6}}-n_{(\bar 3,2)_{-1/6}}&=M_{\boldsymbol{10}}\,,\\
          n_{(\bar 3,1)_{-2/3}}-n_{(3,1)_{+2/3}}&=M_{\boldsymbol{10}}-N_Y\,,\\
          n_{(1,1)_{+1}}-n_{(1,1)_{-1}}&=M_{\boldsymbol{10}}+N_Y\,.
          \label{equsplit10}\end{split}
      \end{align}

     \begin{table}[t]
        \center
        $\begin{array}{c|ccc}
          \hline
          \text{Curve} & SU(5)_\perp-\text{Charge} & \text{Equation} & \text{Homology} \\
          \hline
          \mathbf{5}_{H_u}  & -2t_1    & a_6 a_7 a_8 + a_5 a_7 a_9 + a_4 a_8 a_9 & -c_1 + \tilde{\chi}        \\
          \mathbf{5}_{1}    & -t_1-t_3 & a_1 - c a_4 a_6 a_8 - c a_4 a_5 a_9     & \eta - 2c_1 - \tilde{\chi} \\
          \mathbf{5}_{2}    & -t_1-t_4 & a_1 - c a_5 a_6 a_7 - c a_4 a_5 a_9     & \eta - 2c_1 - \tilde{\chi} \\
          \mathbf{5}_{3}    & -t_1-t_5 & a_1 - c a_5 a_6 a_7 - c a_4 a_6 a_8     & \eta - 2c_1 - \tilde{\chi} \\
          \mathbf{5}_{4}    & -t_3-t_4 & a_5 a_7 + a_4 a_8                       & -c_1 + \chi_7 + \chi_8     \\
          \mathbf{5}_{5}    & -t_3-t_5 & a_6 a_7 + a_4 a_9                       & -c_1 + \chi_7 + \chi_9     \\
          \mathbf{5}_{6}    & -t_4-t_5 & a_6 a_8 + a_5 a_9                       & -c_1 + \chi_8 + \chi_9     \\
          \mathbf{10}_{1}   & t_1      & a_1                                     & \eta - 2c_1 -\tilde{\chi}  \\
          \mathbf{10}_{2}   & t_3      & a_4                                     & -c_1 + \chi_7              \\
          \mathbf{10}_{3}   & t_4      & a_5                                     & -c_1 + \chi_8              \\
          \mathbf{10}_{4}   & t_5      & a_6                                     & -c_1 + \chi_9              \\
          \hline
        \end{array}$
        \caption{Matter curves and their homology classes.}
        \label{tab:a}
      \end{table}      
      As was already mentioned in the introduction to this section, the $U(1)$ fluxes are turned on along
      the branes which intersect $S_{GUT}$ and thus cannot be determined even in the semilocal approach.
      We are therefore allowed to treat the $M$'s as free parameters up to two constraints: 
      \begin{align}
        \label{equ:constraintsum}
        \sum M_{\mathbf{10}}+\sum M_{\mathbf{5}}&=0\,, &
        M_{\mathbf{10}_1}&=-(M_{\mathbf{5}_1}+M_{\mathbf{5}_2}+M_{\mathbf{5}_3})\,.
      \end{align}
      The first equation follows from the tracelessness condition for the four $U(1)$ fluxes
      $\sum_iF_{U(1)_i}=0$ and implies anomaly cancellation (see also \cite{Marsano:2010sq}).   
      The second one holds because if one defines the flux restrictions of the three remaining
      independent $U(1)$ fluxes to the $\mathbf 5$ curves $\mathbf 5_1$, $\mathbf 5_2$ and $\mathbf 5_3$
      to be $M_{-t_1-t_3}=M_{\mathbf{5}_1}$, $M_{-t_1-t_4}=M_{\mathbf{5}_2}$ and
      $M_{-t_1-t_5}=M_{\mathbf{5}_3}$, we can express the $U(1)$ flux restrictions to $\mathbf{10}_1$ as
      $M_{\boldsymbol{10}_1}=M_{t_1}=M_{-(-3t_1+2t_1)}=M_{-(-3t_1-t_3-t_4-t_5)}=M_{-(-t_1-t_3-t_1-t_4-t_1-t_5)}
      =-(M_{\mathbf{5}_1}+M_{\mathbf{5}_2}+M_{\mathbf{5}_3})$.

      For the hypercharge flux, as anticipated earlier, there are more constraints because one must
      prevent it from receiving a Green--Schwarz mass, which is only possible if the flux is switched on
      exclusively along 2-cycles in $S_{GUT}$ which are homologically trivial as two-cycles in the CY
      fourfold. This requirement leads to the constraints 
      \begin{align}
        F_Y \cdot c_1 &= 0 \,,& F_Y \cdot \eta &= 0\,.
      \end{align}
      It is interesting to note that from this it follows that
      \begin{align}
        \sum_{\boldsymbol{5}} N=\sum_{\boldsymbol{10}}N=0\,,
      \end{align}
      so the fields in the representations $n_{(1,2)_{+1/2}}$, $n_{(\bar 3,1)_{-2/3}}$ and
      $n_{(1,1)_{+1}}$ have no net chirality. 
      
      Regarding the column of Table~\ref{tab:a} that displays the homology classes, we see that the
      hypercharge restrictions to the curves are determined solely by $N_7$, $N_8$ and $N_9$. The final
      values for $N_Y$ and $M$, needed for the calculation of concrete spectra that will be performed in
      the next section, is shown in Table~\ref{tab:restrictions}, where $\widetilde N=N_7+N_8+N_9$. 
      
      \begin{table}
        \centering$\begin{array}{c|cc}
          & N_Y& M\\[1mm]
          \hline
          &&\\
          \text{\textbf{10 Curves}}&&\\
          \mathbf{10}_1 &  -\widetilde N &-(M_{\mathbf{5}_1}+M_{\mathbf{5}_2}+M_{\mathbf{5}_3}) \\
          \mathbf{10}_2 &  N_7 &M_{\mathbf{10}_2} \\
          \mathbf{10}_3 & N_8 &M_{\mathbf{10}_3} \\
          \mathbf{10}_4 &  N_9 &M_{\mathbf{10}_4}  \\
          &&\\ 
          \text{\textbf{5 Curves}} & &\\
          \mathbf{5}_{H_u}  & \widetilde N & M_{\mathbf{5}_{H_u}}\\
          \mathbf{5}_{1}  &-\widetilde N &M_{\mathbf{5}_1}\\
          \mathbf{5}_{2} & -\widetilde N&M_{\mathbf{5}_2}\\
          \mathbf{5}_{3}  & -\widetilde N&M_{\mathbf{5}_3}\\
          \mathbf{5}_{4}  & N_7+N_8 &M_{\mathbf{5}_4}\\
          \mathbf{5}_{5}  & N_7+N_9&M_{\mathbf{5}_5}\\
          \mathbf{5}_{6}  & N_8+N_9&M_{\mathbf{5}_6}\\
        \end{array}$
        \caption{Restrictions of hypercharge and $U(1)$ fluxes to the curves. \label{tab:restrictions}}
      \end{table}

      It is important to note that a split of the up-type Higgs curve inevitably leads to a split
      of the $\mathbf{10}_1$ curve. 

    \subsection{Semilocal Embedding of Case I\label{subsec:covercalculation}}
      Using the above spectral cover results and the setup established so far,
      we will now start to calculate the concrete spectrum.      
      The $\mathbf{10}$ curves that accommodate SM matter are fixed to be $\mathbf{10}_1$ and
      $\mathbf{10}_3$. We require that there are three net $\boldsymbol{10}$'s after splitting the curves.
      This leads to the requirements that
      \begin{align}
        M_{\mathbf{10}_1}+M_{\mathbf{10}_3}&=3\,, & N_7+N_9&=0\,.
      \end{align}
      Furthermore, we require  $M_{\mathbf{10}_1}\geq 1$ and $M_{\mathbf{10}_1}+N_8\geq 0$ to have a heavy
      top quark.  Similarly, we find for
      the $\ol[.5]{\boldsymbol{5}}$ curves that
      \begin{align}
        M_{\mathbf{5}_3}+M_{\mathbf{5}_6}&=-3\,, & N_7&=0\,.
      \end{align}
      Hence, also $N_9=0$, and the only remaining parameter that can be used to split some curves is $N_8$.

      Let us continue with the other two $\mathbf{10}$ curves
      that are not associated with SM matter and should therefore better have no net chirality. Using
      Table~\ref{tab:restrictions}, one sees that this can be achieved easily by simply setting 
      \begin{align}
        M_{\mathbf{10}_2}=M_{\mathbf{10}_4}=0\,.
      \end{align}
      Since we do not want zero modes from the $\mathbf{5}_{5}$ curve, we also set $M_{\mathbf{5}_5}=0$.  Thus, we can
      achieve a satisfactory matter sector. 
      
      \begin{table}
        \center$
        \begin{array}{c|cc}
          & n_{(3,1)_{-1/3}}-n_{(\bar 3,1)_{+1/3}}& n_{(1,2)_{+1/2}}-n_{(1,2)_{-1/2}} \\[1mm]
          \hline
          \mathbf{5}_{H_u}& M_{\mathbf{5}_{H_u}} &M_{\mathbf{5}_{H_u}}+N_8 \\
          \mathbf{5}_{1} &  M_{\mathbf{5}_{1}} &M_{\mathbf{5}_{1}}-N_8 \\
          \mathbf{5}_{2} & M_{\mathbf{5}_{2}} &M_{\mathbf{5}_{2}}-N_8 \\
          \mathbf{5}_{4} &  M_{\mathbf{5}_{4}} &M_{\mathbf{5}_{4}}+N_8  \\
        \end{array}
        $
        \caption{Chiralities in terms of $U(1)$ and hypercharge flux restrictions for the Higgs-like $\mathbf
          5$ curves after imposing the matter sector constraints for case I. \label{tab:chir5res}}  
      \end{table}
      
      We now turn to the Higgses. The chiralities of
      the Higgs-like $\mathbf{5}$ curves are shown in Table~\ref{tab:chir5res} in terms of the $U(1)$ and
      hypercharge flux restrictions.  The $M$'s are constrained by the condition~(\ref{equ:constraintsum})
      which now reads
      \begin{align}
        M_{\boldsymbol{5}_{H_u}}+M_{\boldsymbol{5}_1}+M_{\boldsymbol{5}_2}+M_{\boldsymbol{5}_4} &=0\,. 
      \end{align}
      The split of the Higgs-like  $\mathbf{5}$ curves 
      is controlled by the parameter $N_8$, as was already noted above. We need one down-type
      Higgs doublet and one up-type Higgs doublet. The easiest way to realise this is if the down-type
      Higgs doublet, being located on a $\ol[.5]{\mathbf{5}}$, has a chirality of $-1$ whereas the up-type
      Higgs doublet has a chirality of $+1$. Neither the down-type Higgs nor the up-type Higgs must have a
      light triplet. Since we always assume that fields which appear in vector-like pairs become massive,
      the simplest way to get rid of the triplets would be to set the corresponding $M$'s to zero.  
      
      However, we can possibly tolerate triplets if they become heavy with VEVs switched on. Giving a VEV
      to the singlet $\mathbf{1}_2$, there are two allowed couplings between the four $\mathbf{5}$ curves
      of interest:
      \begin{align}
        \label{equ:couplings}
        \ol[.5]{\mathbf {5}}_1\mathbf{5}_4\mathbf{1}_2\,,\\
        \label{equ:couplings2}
        \mathbf{5}_{H_u}\ol[.5]{\mathbf {5}}_2\mathbf{1}_2\,.
      \end{align}
      Hence we can pairwise decouple a triplet from the $\mathbf {5}_{H_u}$ with an antitriplet from the
      $\mathbf {5}_2$ curve. 
      At the same time, we do not want to decouple the up-type Higgs doublet.
      Formulating these requirements in terms of the flux restrictions, we have for the up-type Higgs
      \begin{align}
        \label{equ:up}
        \begin{aligned}
          \text{doublets:}&&M_{\mathbf 5_{H_u}}+N_8+(M_{\mathbf{5}_{2}}-N_8)&=1\,,\\ 
          \text{triplets:}&&M_{\mathbf 5_{H_u}}+M_{\mathbf 5_{2}}&=0\,,
        \end{aligned}
      \end{align}
      which is a contradiction. Therefore, one generically gets a spectrum which contains additional
      exotic fields \cite{Marsano:2009gv}, or no up-type Higgs.  Note that, since both the $\boldsymbol{5}_{H_u}$ and
      the $\boldsymbol{5}_2$  
      are split with $\widetilde{N}$, this result is independent of the hypercharge flux.
      
      A loophole would be to choose $\ol[.5]{\mathbf{5}}_{2}$ as the down-type Higgs curve. Then the spectrum is
      free of exotics, but the coupling~(\ref{equ:couplings2}) is nothing but a $\mu$-term. Furthermore, as
      discussed before, the $\ol[.5]{\mathbf{5}}_2$ curve has no $t_3$ factor, so there are no down-type masses in this
      model.  One can nevertheless realise this spectrum with the parameter choice
      \begin{align}
        \label{equ:firstchoice}
        M_{\mathbf{5}_{H_u}}=M_{\mathbf{5}_{1}}=M_{\mathbf{5}_{2}}=M_{\mathbf{5}_{4}}=0\,,\quad N_8=1\,.
      \end{align}
      The remaining doublets from $\boldsymbol{5}_1$ and $\boldsymbol{5}_4$ can be decoupled by the
      term~(\ref{equ:couplings}). 
      
      If we want to embed the model from Section~\ref{subsec:higgs}, which has $\boldsymbol{5}_4$ as the
      down-type Higgs, we can use a similar argument as in Eq.~(\ref{equ:up}) to arrive at the mutually
      contradicting constraints
      \begin{align}
        \begin{aligned}
          \text{doublets:}&& M_{\mathbf 5_{4}}+N_8+(M_{\mathbf{5}_{1}}-N_8)=-1\,,\\
          \text{triplets:}&&M_{\mathbf 5_{4}}+M_{\mathbf 5_{1}}=0\,.
        \end{aligned}
      \end{align}
      An example spectrum one can get using $\ol[.5]{\mathbf{5}}_{4}$ as the down-type Higgs curve is shown in
      Table~\ref{tab:chir5-2} with parameters
      \begin{align}
        \label{equ:secondchoice}
        M_{\mathbf{5}_{H_u}}=0, \quad M_{\mathbf{5}_{1}}=M_{\mathbf{5}_{2}}=1\,, \quad
        M_{\mathbf{5}_{4}}=-2,\quad N_8=1\,. 
      \end{align}
      $\mathbf{5}_{4}$ has the desired doublet but also two triplets, one of which can be combined
      with the triplet of the $\mathbf{5}_{1}$ via the coupling
      $\ol[.5]{\mathbf{5}}_1\mathbf{5}_4\mathbf{1}_2$. The other one, however, will remain light and apart
      from that there is another unwanted triplet in the $\mathbf{5}_{2}$.
      
      \begin{table}
        \center
        \subtable[Parameter Choice~(\ref{equ:firstchoice})\label{tab:chir5-1}]{
          $\begin{array}{c|cc}
            &  \text{Triplets}&\text{Doublets}\\[1mm]
            \hline
            \mathbf{5}_{5}&0&0\\
            \mathbf{5}_{H_u}&0 &1 \\
            \mathbf{5}_{1} &0 &-1 \\
            \mathbf{5}_{2} &0 &-1 \\
            \mathbf{5}_{4} &0 &1  \\
          \end{array}$}
        \hspace{2cm}
        \subtable[Parameter Choice~(\ref{equ:secondchoice})\label{tab:chir5-2}]{
          
          $\begin{array}{c|cc}
            &   \text{Triplets} & \text{Doublets}\\[1mm]
            \hline
            \mathbf{5}_{5}&0&0\\
            \mathbf{5}_{H_u}&0 &1 \\
            \mathbf{5}_{1} &1 &0 \\
            \mathbf{5}_{2} &1 &0 \\
            \mathbf{5}_{4} &-2 &-1  \\
          \end{array}$
        }
        \caption{Two possible splits of the Higgs curves for case~I.}
      \end{table}

    \subsection{Semilocal Embedding of Case II}
      Since there is only one $\mathbf{10}$ curve in this model that can carry SM matter, we have to require 
      \begin{align}
        M_{\mathbf{10}_1}=3,\quad N_{\mathbf{10}_1}=-\widetilde N=0.
      \end{align}
      The first condition is unproblematic because it implies 
      \begin{align}
        M_{\mathbf{5}_1}+M_{\mathbf{5}_2}+M_{\mathbf{5}_3}=-3,
      \end{align}
      which is exactly what we need since $\ol[.5]{\mathbf{5}}_1$, $\ol[.5]{\mathbf{5}}_2$ and $\ol[.5]{\mathbf{5}}_3$
      are the possible 
      matter curves. Furthermore, $\widetilde{N}=0$ implies that the matter curves are not split, so we end up with a
      reasonable matter sector. On the other hand, $\widetilde{N}=0$ inhibits a split of the up-type Higgs because
      $N_{H_u}=\widetilde N$, see 
      Table~\ref{tab:restrictions}. There is also no way of coupling the up-type Higgs curve to some other even matter parity
      Higgs curve to make the triplet heavy because the coupling has the charges $-2t_{\text{even}}+2t_{\text{even}}$ in terms
      of the notation introduced in Section~\ref{sec:IImatter}, which cannot be cancelled with even matter parity
      singlets.  Therefore, we can conclude that also in case II it is not possible to arrive at a satisfying spectrum
      while giving masses to quarks and leptons.
      
      \medskip
      
      Note that because of the way the hypercharge flux restricts to the Higgs-like curves, it is not possible to achieve a
      satisfactory Higgs sector. This is true in both matter parity cases, and even when allowing for exotics from the matter
      sector. Hence, it is ultimately the problem of doublet-triplet splitting which prohibits a semilocal embedding of the
      local models.
      

  \section{Conclusions and Outlook\label{sec:conclusion}}
    The incorporation of GUTs in string theory aims at a consistent 
    ultraviolet completion of all fundamental interactions including 
    gravity. Such a consistency can only be assured if we have a 
    consistent global string theory construction. One might still ask 
    the question whether there are some properties of particle 
    physics that could be studied in a bottom-up approach at a local 
    level. In F-theory such a local description concerns a 
    d=4 spacetime (point in extra dimensions) or a semilocal description 
    (d=8 spacetime with four extra dimensions). Local descriptions give more 
    freedom for model building but might not have valid ultraviolet 
    completions and could thus be inconsistent. 
    
    In the present paper we have analysed the 
    local $E_8$ point for the construction of an $SU(5)$ GUT 
    and identified exactly two models that are consistent with 
    sufficient proton stability and nontrivial masses for all 
    quarks and leptons. 
    They might be candidates for a realistic string version 
    of the MSSM, although some aspects (such as Majorana neutrino masses) 
    have not yet been addressed. It is interesting to see that proton 
    stability can be implemented at the local point.
    This is in contrast to results in the 
    heterotic theory where such mechanisms required some amount of 
    nonlocality within the known consistent global constructions\cite{Forste:2010pf,Lee:2010gv}. 
    
    Unfortunately the two local models mentioned above do not 
    possess a consistent global completion. 
    Our analysis using spectral cover techniques proves the 
    inconsistency of the otherwise acceptable local models
    even at the level of the semilocal construction. 
    This is one of the central results of our analysis. 
    Other studies of the $E_8$ point \cite{Dudas:2009hu,Marsano:2009wr,King:2010mq,Choi:2010nf,Pawelczyk:2010xh} have
    assumed that the  
    existence of $P_M$ requires a nonlocal mechanism. This implies 
    that a crucial aspect of the MSSM is not provided 
    by the local point and thus undermines its predictive power. 
    
    Recently, it has been argued\cite{Cecotti:2010bp} that the spectral cover 
    considerations might possibly not be the only way to include 
    semilocal effects\footnote{Note added: While finishing this work, the paper\cite{Chiou:2011js} appeared which
      discusses Yukawa  couplings in the T-brane setup.}. It might be 
    interesting to see whether  more general approaches could validate the two models we have 
    identified (maybe even without the need for nonlocalities). 
    More work in this direction is surely needed. Still it 
    seems that there is no alternative to global consistent 
    model building. We cannot trust the predictions of local
    models as long as they are not confirmed by global 
    constructions. In a more positive interpretation this 
    tells us that string theory is more than just "bottom-up"
    model building and that we can learn nontrivial things for particle physics
    from the full string theory.

  \section*{Acknowledgements}
    We are grateful to Thomas Grimm for helpful discussions.
    This work was partially supported by the SFB-Transregio TR33 "The Dark Universe" (Deutsche
    Forschungsgemeinschaft) and the European Union 7th network program "Unification in the LHC era"
    (PITN-GA-2009-237920). 

    
  \newpage

  \appendix
  \section{$\ol[.5]{\boldsymbol{5}}_1$ as Down-Type Higgs\label{app:higgs}}
    Choosing the curve $\ol[.5]{\mathbf{5}}_{1}$ as the down-type Higgs curve, we have the gauge invariant
    couplings 
    \begin{align}
      \ol[.5]{\mathbf{5}}_{H_d}\mathbf{10}_1\ol[.5]{\mathbf{5}}_6, \quad
      \ol[.5]{\mathbf{5}}_{H_d}\mathbf{10}_3\ol[.5]{\mathbf{5}}_3,
    \end{align}
    which lead to a rank-two down-type Yukawa matrix if the curves are not split and three generations
    come from two curves. We now ask whether the curves can be split in a way such that the rank is
    reduced to zero or one. 
    
    The relevant couplings in $\ol[.5]{\mathbf{5}}_{H_d}\mathbf{10}_M\ol[.5]{\mathbf{5}}_M$ in terms of SM
    representations are the ones which involve the Higgs doublet $D$: 
    \begin{align} 
      D\ol e L, \quad D\ol d Q
    \end{align}
    The chiralities of the fields are given by the formulae~(\ref{equsplit5}),
    \begin{align}
      \begin{split}
        n_{(3,1)_{-1/3}}-n_{(\bar 3,1)_{+1/3}}&=M_{\boldsymbol{5}}\\
        n_{(1,2)_{+1/2}}-n_{(1,2)_{-1/2}}&=M_{\boldsymbol{5}}+N_Y
      \end{split}
    \end{align}
    for the $\mathbf{5}$ curves and~(\ref{equsplit10}) for the $\mathbf{10}$ curves:
    \begin{align}
      \begin{split} 
        n_{(3,2)_{+1/6}}-n_{(\bar 3,2)_{-1/6}}&=M_{\boldsymbol{10}}\\
        n_{(\bar 3,1)_{-2/3}}-n_{(3,1)_{+2/3}}&=M_{\boldsymbol{10}}-N_Y\\
        n_{(1,1)_{+1}}-n_{(1,1)_{-1}}&=M_{\boldsymbol{10}}+N_Y
      \end{split}
    \end{align}
    
    Our primary concern is the quark Yukawa matrix. Since the anti-down quark belongs to the
    representation $n_{(\bar 3,1)_{+1/3}}$ and the down quark belongs to $n_{(3,2)_{+1/6}}$, their
    chiralities are in our case fully determined by $M_{\boldsymbol{5}_3}$, $M_{\boldsymbol{5}_6}$ and
    $M_{\boldsymbol{10}_1}$ and $M_{\boldsymbol{10}_3}$. Overall, we need three generations from the
    $\ol[.5]{\mathbf{5}}$ matter curves and three 
    generations from the $\mathbf{10}$ matter curves and therefore we have the conditions 
    \begin{align}
      M_{\boldsymbol{5}_3}+M_{\boldsymbol{5}_6}=-3, \quad
      M_{\boldsymbol{10}_1}+M_{\boldsymbol{10}_3}=3.
    \end{align}
    
    If one sets one of the $M_{\boldsymbol{10}}$'s equal to one and the other one equal to two, we
    arrive at nothing new. Explicitly, choosing $M_{\boldsymbol{10}_1}=1$ and $M_{\boldsymbol{10}_3}=2$,
    one would demand that a heavy bottom quark is generated through the coupling
    $\ol[.5]{\mathbf{5}}_{H_d}\mathbf{10}_1\ol[.5]{\mathbf{5}}_6$ and thus 
    set $M_{\boldsymbol{5}_6}=-1$ and $M_{\boldsymbol{5}_3}=-2$. But then the other term,
    $\ol[.5]{\mathbf{5}}_{H_d}\mathbf{10}_1\ol[.5]{\mathbf{5}}_6$, also exists and the matrix has rank two,
    which is exactly what we had before. 
    
    Thus, the only new option is $M_{\boldsymbol{10}_1}=3$ and $M_{\boldsymbol{10}_3}=0$ where the
    remaining relevant coupling is $\ol[.5]{\mathbf{5}}_{H_d}\mathbf{10}_1\ol[.5]{\mathbf{5}}_6$.
    $M_{\boldsymbol{5}_6}=0$ then leads to a rank-zero matrix and all other values for
    $M_{\boldsymbol{5}_6}$ yield a rank-one matrix. In order not to introduce unnecessary fields, that
    is chiralities larger than three, one can see from the formulae for the other representations $n_{(\bar
      3,1)_{-2/3}}-n_{(3,1)_{+2/3}}=M_{\boldsymbol{10}}-N_Y$ and
    $n_{(1,1)_{+1}}-n_{(1,1)_{-1}}=M_{\boldsymbol{10}}+N_Y$ that in the rank-one or -zero case it
    is in addition necessary that $N_{\boldsymbol{10}_1}=N_{\boldsymbol{10}_3}=0$. Thus, this solution
    is a rather trivial one. It is important to note that nowhere in the above argumentation any use was made of the homology
    classes and the corresponding correlations between the different $M$'s and $N$'s as determined by
    the spectral cover approach.

\end{document}